# The Optical Gravitational Lensing Experiment. Deep Photometry of the Sagittarius Dwarf Spheroidal Galaxy*

Mario Mateo[1],
e-mail I: mateo@astro.lsa.umich.edu

A. Udalski[2], M. Szymański[2],
e-mail I: (udalski,msz)@sirius.astrouw.edu.pl

J. Kałużny[2], M. Kubiak[2], W. Krzemiński[3]
e-mail I: (jka,mk)@sirius.astrouw.edu.pl, wojtek@roses.ctio.noao.edu

## ABSTRACT

We present deep CCD photometry of a field in the newly discovered dwarf spheroidal galaxy in Sagittarius (hereafter Sgr), and of a nearby control field. These data (totaling ∼3.5 hrs exposure time in V and I) were used to produce color-magnitude (CM) diagrams reaching I ∼ 22.3 in both fields. After statistically removing the field-stars from the CM diagram we find that Sgr is dominated by a moderately old (age ∼ 10 Gyr) population, significantly younger than a typical globular-cluster population. There is some evidence for a weak intermediate-age component or alternatively a population of blue stragglers. These results confirm that Sgr is a *bona fide* dwarf spheroidal (dSph) galaxy, the ninth found orbiting the Milky Way. We have discovered nine short-period variables in the Sgr field, well in excess of the number found in the control field. Seven of these stars have similar apparent magnitudes and are almost certainly RR Lyr stars in Sgr. We have used the minimum-light colors of the RRab variables to determine the reddening of the Sgr field to be E(V−I) = 0.22. For an assumed RR Lyr luminosity of $M_V = +0.6$ we conclude that the distance of Sgr is $25.2 \pm 2.8$ kpc. The properties of the Sgr giant branch and upper main sequence are consistent with a mean metallicity of [Fe/H] $\sim -1.1 \pm 0.3$.

---

*Based on observations obtained at the Las Campanas Observatory of the Carnegie Institution of Washington

[1]Department of Astronomy, University of Michigan, 821 Dennison Bldg., Ann Arbor, MI 48109–1090

[2]Warsaw University Observatory, Al. Ujazdowskie 4, 00–478 Warszawa, Poland

[3]Carnegie Observatories, Las Campanas Observatory, Casilla 601, La Serena, Chile

## 1. Introduction

The Optical Gravitational Lensing Experiment (OGLE) is a long term project with the main goal of searching for dark matter in our Galaxy by identifying microlensing events towards the Galactic Bulge (Paczyński 1986; Udalski *et al.* 1992). At times the Bulge is either unobservable or conditions are too poor to effectively monitor stars in these very dense fields, and we then turn to other programs that do not demand such excellent observing conditions but still greatly benefit from long-term photometric observations. This *Letter* describes the first results of one such project to obtain deep, multi-epoch photometry of a newly-discovered dwarf galaxy near the Galactic center.

Ibata, Gilmore and Irwin (1994; hereafter IGI) announced the discovery of the new galaxy during the course of a large-scale kinematic survey of the Galactic Bulge. Because the galaxy is located in the constellation Sagittarius, IGI christened it the Sgr dwarf. This naming follows convention and the authors' suspicion that Sgr is probably a dwarf spheroidal (dSph) system. IGI presented a photographic BR photometry of the galaxy, which reveals clear evidence for an old or intermediate-age population superimposed on a rich population of mostly foreground disk and Bulge stars. These data were not deep enough to allow IGI to determine the age of the predominant population in Sgr directly. Moreover, the extreme contamination by Galactic field stars makes any photographic study of main sequence stars in Sgr virtually impossible.

IGI noted that the distribution of stars in the upper giant branch and red horizontal branch (RHB) is similar to the intermediate-age populations of the Small Magellanic Cloud (Gardiner and Hatzidimitriou 1992). They also identified a number of Carbon star candidates in the Sgr field. In these respe¡cts, Sgr is similar to the Fornax dSph galaxy (Buonanno *et al.* 1985). The mean metallicity and total luminosity of Sgr are also comparable to those of Fornax, and, like all other dSph galaxies, Sgr seems to lack any significant HI emission. Only the complex structure of Sgr bears any resemblance to a dwarf irregular galaxy, but even this does not discount the possibility that it is a dSph. Sgr is located very near the Galactic center (IGI concluded that the galactocentric distance of Sgr is 16 kpc, by far the closest external galaxy), and tidal forces could readily account for Sgr's distorted structure (*e.g.*, see Freeman 1993).

In this *Letter* we present deep CCD photometry of Sgr which strongly supports the identification of this galaxy as a dSph system. These data provide the first direct determination of the age of bulk of the stellar population in Sgr. In addition, we use the properties of newly-discovered RR Lyr stars in Sgr to obtain a more precise estimate of its reddening and distance.

## 2. Observations

The OGLE microlensing search is conducted using the 1-m Swope telescope at Las Campanas Observatory which is operated by Carnegie Institutions of Washington. A single $2048 \times 2048$ pixels Loral CCD chip, giving the scale of 0.44 arcsec/pixel at 1-m telescope focus, is used as the detector. New observations are fed into a data-reduction pipeline which processes the CCD images in near-real time. Details of the standard OGLE reduction techniques are described by Udalski *et al.* (1992).

The new data for the Sgr galaxy were obtained on nearly every night from 1-17 July 1994 (UT). These data were nearly always obtained as VI pairs (the I-band filter matches the Kron-Cousins I-band), with individual exposure times ranging from 500-1200 seconds. Our primary Sgr field is located at $\alpha_{2000} = 19^h06^m11^s$, and $\delta_{2000} = -30°25'22''$; $(l, b) = (5°, -15°)$. We also monitored a control field at $\alpha_{2000} = 18^h33^m31^s$, and $\delta_{2000} = -41°10'41''$, which corresponds to the reflection of the Sgr field about $l = 0°$ (*i.e.*, $(l, b) = (-5°, -15°)$). The Sgr field we observed corresponds to the easternmost high-density region identified by IGI. We specifically avoided the highest-density region in Sgr because of its proximity to the globular cluster M 54.

For the present study we produced deep V and I images of the Sgr and control fields as follows. First, we registered the best VI images of each field to common coordinate systems[5]. Only full-pixel shifts were applied to the individual frames; no interpolation of any sort was used. Second, we *summed* the individual exposures to make the final deep images. The total exposures times in each field and in each filter was about 3.5 hrs. A summary of the properties of the summed images is given in Table 1.

The stellar photometry was derived from these deep images using the DoPHOT photometry program (Schechter, Mateo and Saha 1993). Stars with excessively large internal photometric errors compared to the mean error of other stars of similar brightness were weeded out of the results produced by DoPHOT. Only about 5% of all stars identified by DoPHOT were removed during this rejection step.

The data were calibrated during the photometric night of 12 July 1994. Rather than observe standards (which would have taken time away from the primary OGLE observations), we used observations of previously calibrated stars in two of the OGLE fields:

---

[5]In this case, 'best' means frames with a seeing FWHM $\leq 1.8$ arcsec. Sgr was usually observed when the Bulge could not be monitored; that is, during mediocre seeing conditions or at high airmass. The summed frames therefore do not fairly represent what could be achieved with the same exposure time for data obtained under more favorable conditions

BW 6 and BW 7 (see Udalski *et al.* 1992 for a description of these fields). Each field was observed 3-4 times over a wide airmass range. At least 75 well-measured stars spanning a large color range were used to determine the transformation and extinction coefficients for the night and the results were used to calibrate stars in the Sgr and control fields observed that night. These stars then served to calibrate the photometry for stars measured on the summed CCD frames. We estimate that the $1\sigma$ errors in the adopted zero-points are 0.03 mag in I and 0.04 mag in (V−I).

### 3. The Deep Color-Magnitude Diagram of Sgr

Figure 1 shows the calibrated VI color-magnitude (CM) diagrams of the Sgr and control fields. The most obvious feature in both diagrams is the broad swath of stars scattered diagonally from the lower-right to near the center of the diagram and the more-or-less vertical distribution above that (I $\lesssim$ 18). These stars presumably correspond to the Bulge and disk population along the lines of sight. There is clearly an additional population present in the Sgr field that is absent from the control CM diagram; two features in particular stand out. First there is a clump of stars at ((V−I), I) = (1.1, 17) that has no counterpart in the control field CM diagram. Second, there is a large buildup of stars in the Sgr field at I $\gtrsim$ 19.5 and (V−I) $\sim$ 0.8. Some additional subtle features are also visible in the Sgr field CM diagram, such as a line of stars connecting the bright and faint clumps, and possibly a sequence of stars extending upward and redward from the brighter clump. All of these features presumably belong to Sgr itself.

In order to separate the Sgr stars from contaminating field stars, we have devised the following statistical field-subtraction procedure. We first divided the Sgr and control CM diagrams into $0.1 \times 0.25$ mag cells (the smaller dimension is along the color axis) and counted stars in each cell of both diagrams. We then normalized the star counts in the two fields by the number of stars located within the dashed boxes shown in Figure 1. For each cell we then subtracted the corrected number of stars in the control CM diagram from the number in the Sgr CM diagram. Figure 2 shows the resulting diagram which consists primarily (but statistically) of Sgr stars. The sizes of each point in Figure 2 are proportional to $\log_{10}(N_i)$, where $N_i$ is the net number of stars in the $i$th cell after subtraction of the control field. For each cell, $\sigma_N = (N_{Sgr} + N_{control})^{1/2}$, where $N_{Sgr}$ and $N_{control}$ are the counts in the Sgr and control CM diagrams, respectively. Only cells with more than $1.5\sigma_N$ or 4 stars (whichever was larger) are plotted in Figure 2. The mean formal errors in I and (V−I) are also plotted in Figure 2 for I $\geq$ 19.0. For the brighter cells (I $\leq$ 16.5), the results shown in Figure 2 were derived from single VI CCD pairs of both fields. For I $\leq$ 16.5 and (V−I) $\geq$ 1.4 the *individual* stars measured from the same shallow CM diagram are plotted in the Figure 2; that is, we did not perform any field-star subtraction in this part of the diagram. Because of the statistical nature of Figure 2, any single point in the diagram may be the result of bad luck or unfortunate binning. However, any sequence or clump of points is probably significant.

The field-subtracted CM diagram in Figure 2 now shows clearly many of the features that belong exclusively to the Sgr population. The brighter of the two clumps visible in the upper part of Figure 1 is revealed as the red horizontal branch of Sgr. The fainter clump corresponds to unevolved main sequence stars, and the two clumps are clearly connected by a well-defined subgiant branch. In addition, there is a hint of stars blueward of the RHB and of some stars extending above the obvious tip of the main sequence at I $\sim$ 19.5. Many of the isolated cells with net positive counts ((V−I) $\gtrsim$ 1.6 and along the location of the main sequence of the field stars seen in Figure 1) probably result from slight differences in completeness and reddening between the Sgr and control fields.

The top panel of Figure 3 shows the V and I luminosity functions (LFs) derived from the field-subtracted CM diagram of Sgr. These LFs allow us to constrain some of the global photometric properties of Sgr. For example, the total number of Sgr stars we measured to I $\sim$ 22 is about 8100, of which 550 are brighter than the main sequence turnoff at I $\sim$ 19.5. This corresponds to densities of 33.6 and 2.4 stars/arcmin$^2$ for the main sequence and evolved stars, respectively. The integrated apparent I magnitude and (V−I) color of the Sgr stars we measured in our field is 10.3, and 0.8, respectively, while the mean V-band surface brightness ($\Sigma_V$) is 25.8. We extrapolated the LFs in Figure 3 to $M_V = M_I = +13.5$ by adopting the average LF slope in the ranges 21 < V < 22 and 20.25 < I < 21.25 (our star counts are still reasonably complete in these intervals where unevolved main-sequence stare predominate). The corrected surface brightness and color are $\Sigma_V$ = 25.3 and (V−I) = 0.7. From the isopleths given by IGI, we estimate that the densest regions in Sgr have $\Sigma_V$ = 24.9 assuming no variations in the underlying stellar population.

## 4. The RR Lyr Stars of Sgr: Reddening and Distance

From measurements of the individual CCD frames of the Sgr and control fields we have discovered a clear excess of variable stars in the Sgr field. A detailed paper describing these results will be published at the end of the 1994 OGLE observing season. The current data, however, are quite adequate to identify most of these variables as RR Lyr stars and to provide well-sampled light curves and pulsation periods. The locations of these stars in the CM diagram of Sgr are shown in Figure 2. It is apparent that most of these stars are located on the blue extension of the RHB and are almost certainly members of Sgr. Two of the stars are considerably brighter than the others; these may be foreground Galactic RR Lyr stars or brighter variables in Sgr (*e.g.*, anomalous Cepheids). The survey of RR Lyr stars in fields near the Galactic Center by Wesselink (1987) suggests that we would expect 0.5 Galactic RR Lyr stars in our CCD fields. Finding one or two is obviously not inconsistent with this prediction.

The RR Lyr stars are valuable for two reasons: they can be used to determine the reddening in the field and they provide a way to determine a precise distance for Sgr. The former step is possible because of the well-known property that the minimum-light colors of ab-type RR Lyr stars are nearly identical regardless of metallicity (Sturch 1966; Lub 1979; Wesselink 1987) That this holds for (V−I) is demonstrated in Table 2 which lists unreddened (V−I) colors for a number of well-observed field RRab variables (the data in Table 2 come from Liu and Janes 1989). The mean minimum-light (V−I) color is $0.58 \pm 0.03$. Table 2 shows that the minimum-light color appears to be independent of metallicity.

For the five well-observed RRab variables in the Sgr field we obtain $\langle (\mathrm{V-I})_{min} \rangle = 0.80 \pm 0.02$. One of these stars is a likely foreground object, but is sufficiently distant (about 10 kpc) to be behind most of the extinction along this line of sight. From this result we derive $\mathrm{E(V-I)} = 0.22 \pm 0.04$, and a total I-band extinction of $\mathrm{A_I} = 0.33 \pm 0.08$ (Cardelli *et al.* 1989). For a normal extinction law with $\mathrm{R_V} = \mathrm{A_V}/\mathrm{E(B-V)} = 3.1$, we get $\mathrm{E(B-V)} = 0.81\mathrm{E(V-I)} = 0.18 \pm 0.03$, and $\mathrm{A_V} = 1.67\mathrm{A_I} = 0.55 \pm 0.10$. IGI adopted E(B−V) = 0.14 for Sgr based on the Burstein and Heiles (1982) reddening maps.

The average V and I mean magnitudes of the RR Lyr stars that are probable Sgr members are $\langle \mathrm{V} \rangle = 18.16 \pm 0.05$ (seven stars), and $\langle \mathrm{I} \rangle = 17.51 \pm 0.05$ (six stars). Two of these stars are c-type variables. Because of crowding with a neighboring red star, one of the Sgr RR Lyr stars has too few I-band measurements to determine its mean brightness, but its V-band light curve is well defined. If we assume the mean V-band absolute magnitude of RR Lyr stars is $0.6 \pm 0.2$, we derive a true distance modulus of $\mu_0 = 17.01 \pm 0.24$ for Sgr (the uncertainty is dominated by the error in $M_{V,RR}$). This modulus corresponds to a distance of $25.2 \pm 2.8$ kpc, in excellent agreement with the preliminary result ($24 \pm 2$ kpc) derived by IGI.

## 5. Discussion

The lower part of Figure 2 shows the field-subtracted CM diagram of Sgr with two isochrones from VandenBerg (1985) and VandenBerg and Bell (1985). The older isochrone corresponds to an age of 10 Gyr and a metallicity of $[\mathrm{Fe/H}] = -1.2$, while the younger isochrone is for an age of 4 Gyr and [Fe/H] = $-1.0$. Given the errors of the photometry and the binning imposed by the field-subtraction algorithm, it is clear that the 10 Gyr isochrone best fits the bulk of the stars in Sgr. Significantly older isochrones are either too faint or too blue (if a lower metallicity is adopted) to fit the upper main sequence stars in Sgr. We conclude that Sgr definitely contains a substantial population of stars as young as about 10 Gyr. The younger isochrone plotted in Figure 2 could plausibly account for the short extension of stars above the tip of the main sequence with I $\lesssim$ 19.5, implying that Sgr contains stars as young as 4 Gyr. The presence of Carbon stars in Sgr (IGI) is compatible with the existence of an intermediate-age component in the galaxy. Alternatively, these stars may be blue stragglers that are as old as the bulk of the stars in Sgr.

The isochrones shown in Figure 2 imply a relatively high mean metallicity for Sgr; [Fe/H] = $-1.1 \pm 0.3$ is consistent with the fits shown in Figure 2. This result agrees well with the preliminary metallicity determination by IGI based on their photographic CM diagram and their spectroscopy of some Sgr red giants.

Figure 2 also shows three globular cluster giant-branch sequences from the compilation of Da Costa and Armandroff (1990). The sequences range from [Fe/H] = $-0.7$ to $-1.6$; further details are given in the caption. This comparison is of limited utility because of the binning of the CM diagram and the lack of any field-star subtraction in the red giant region of the diagram. Nonetheless, the locations of the more giant-branch sequences of the more metal-rich clusters (Figure 2) seem to be most consistent with the sparse Sgr giant branch observed in our field, in broad agreement with the metallicity estimates from the isochrone comparisons.

We can also constrain the age of Sgr by comparing its luminosity function (LF) with those of other stellar systems. The lower panel of Figure 3 compares the V-band LF with the LFs of the Carina dSph galaxy (Mighell 1990b) and the Galactic globular NGC 7492 (Côté *et al.* 1991). The normalization of these LFs is described in the caption. Carina contains a significant intermediate-age population (Mould and Aaronson 1983; Mighell 1990a), perhaps as young as 5-7 Gyr. The age of NGC 7492 is about 14 Gyr (Côté *et al.* 1991), similar to the canonical age of most Galactic globulars. The point of this comparison is that the LF near the main sequence turnoff (in the range 20.0 $\lesssim$ V $\lesssim$ 21.5 in Figure 3) is an excellent age indicator (Paczyński 1984). The progression of the LFs from left to right (Carina, Sgr and NGC 7492, respectively) confirms that the bulk of the population of Sgr is significantly younger than the canonical age of the globular clusters, though not as young as Carina. This is not incompatible with the isochrone results or the presence of RR Lyr stars in Sgr. Based on studies of SMC clusters, Stryker *et al.* (1985) and Olszewski *et al.* (1987) concluded that RR Lyr stars first become visible in clusters older than about 10 Gyr.

Our results and the results presented by IGI provide overwhelming evidence that Sgr is a *bona fide* dSph galaxy, the ninth known in orbit about the Milky Way. Based on their more extensive (though shallower) photometry, IGI showed that the total luminosity of Sgr is similar to that of the Fornax dSph. In contrast, the stellar population of Sgr is most similar to that of the Sculptor (Da Costa 1984), Draco (Olszewski and Aaronson 1985), and Sextans (Mateo *et al.* 1991) dwarfs (see also Hodge 1989 for a summary of the ages of dSph systems). Each of these systems also shows evidence for a population of stars slightly younger than the stars in globular clusters, and all contain many RR Lyr stars.

In its integrated properties, Sgr shows some interesting differences from other dSph galaxies. van den Bergh (1994) recently noted a possible age-Galactocentric distance relationship for low-luminosity dwarfs near our Galaxy with the oldest systems being located closest to the Milky Way. Sgr seems to mildly violate this relationship. 'Mildly' because although Sgr is by far the closest dSph system to the Galaxy, it is clearly not like the more extreme dSph systems which exhibit strong evidence of a large intermediate-age population (*e.g.*, Leo I, Fornax or Carina). Our confirmation of the relatively high metallicity found by IGI places Sgr squarely on the [Fe/H]-luminosity relation of Caldwell *et al.* (1992; IGI concluded that $M_{V,Sgr} \sim -13$ by comparing the numbers of evolved stars in Fornax and Sgr). However, Sgr lies well off their surface brightness-[Fe/H] relation. We interpret this as additional evidence that Sgr is experiencing extreme tidal disruption and can quantify the extent of this disruption by noting that the V-band central surface brightness of Sgr should be 21.5 if it obeyed the Caldwell *et al.* (1992) relation. The observed value is 24.9 (section 3).

The calibrated photometry of the Sgr and control fields can be obtained via the Internet from host "sirius.astrouw.edu.pl" (148.81.8.1), using the "anonymous ftp" service (directory "ogle/sgr"; files "sgr.cal", "control.cal" and "README"). Information on the recent status of the OGLE project is also available via "World Wide Web" WWW: "http://www.astrouw.edu.pl/".

This project was supported by NSF grants AST 91-18086 and AST 92-23968 MM, AST 92-16494 to Bohdan Paczyński and AST 92-16830 to George Preston, and by Polish KBN grants No 2-1173-9101 and BST475/A/94 to MK. We thank Ed Olszewski, Taft Armandroff, and Nick Suntzeff for their useful comments on an early draft of this paper.

Table 1. Properties of the Summed CCD Images

| Field | Filter | Exposure Time (seconds) | Seeing (arcsec) | $N_{frames}$ |
|---|---|---|---|---|
| Sgr | I | 12463 | 1.50 | 15 |
| Sgr | V | 13800 | 1.50 | 15 |
| | | | | |
| Control | I | 12850 | 1.58 | 15 |
| Control | V | 12900 | 1.45 | 14 |

Table 2. Minimum-Light (V−I) Colors of Field RR Lyr Stars

| Star | (V−I) | E(B−V) | E(V−I) | $(V-I)_0$ | [Fe/H] |
|---|---|---|---|---|---|
| SW And | 0.65 | $0.06 \pm 0.01$ | $0.07 \pm 0.01$ | 0.58 | −0.1 |
| RR Cet | 0.63 | $0.03 \pm 0.01$ | $0.04 \pm 0.01$ | 0.59 | −1.3 |
| SU Dra | 0.62 | $0.01 \pm 0.02$ | $0.01 \pm 0.02$ | 0.61 | −1.6 |
| RX Eri | 0.68 | $0.05 \pm 0.02$ | $0.06 \pm 0.02$ | 0.62 | −1.4 |
| RR Gem | 0.65 | $0.08 \pm 0.02$ | $0.10 \pm 0.02$ | 0.55 | −0.2 |
| RR Leo | 0.63 | $0.05 \pm 0.02$ | $0.06 \pm 0.02$ | 0.57 | −1.2 |
| TT Lyn | 0.63 | $0.01 \pm 0.02$ | $0.01 \pm 0.02$ | 0.62 | −1.4 |
| AV Peg | 0.64 | $0.07 \pm 0.02$ | $0.09 \pm 0.02$ | 0.55 | 0.0 |
| AR Peg | 0.95 | $0.32 \pm 0.04$ | $0.40 \pm 0.05$ | 0.55 | −0.3 |
| TU UMa | 0.61 | $0.02 \pm 0.02$ | $0.02 \pm 0.02$ | 0.59 | −1.3 |
| UU Vir | 0.60 | $0.01 \pm 0.02$ | $0.01 \pm 0.02$ | 0.59 | −0.4 |

Fig. 1.— The calibrated color-magnitude diagrams of our Sgr field (upper) and the control field (lower). The dashed box shows the region that was used to normalize the counts in the two fields prior to field-star subtraction.

Fig. 2.— The field-star subtracted CM diagram of Sgr. In both panels, the sizes of the squares are proportional to $\log_{10}(N_i)$, where $N_i$ is the number of net Sgr stars in each cell. For I $\leq$ 16.5 and (V−I) $\geq$ 1.4, the *individual* stars from a shallow CM diagram are plotted – no field-star removal has been attempted in this part of the diagram. The crosses denote the mean magnitudes and colors of the RR Lyr stars discovered in the Sgr field. In the lower panel, the broad dashed lines correspond to the isochrones described in the text. The dotted lines are the fiducial giant branches of (from right to left) 47 Tuc, NGC 1851, and M 2. These data are from Da Costa and Armandroff (1990).

Fig. 3.— The upper panel plots the V and I differential luminosity functions (LFs) of Sgr after field-star removal. Incompleteness is probably important for I $\gtrsim$ 21.2 and V $\gtrsim$ 22.0. The lower panel compares the V-band differential luminosity functions of Sgr (dark solid line, squares), Carina (dotted line, circles; Mighell 1990b), and NGC 7492 (dashed line, triangles; Côté *et al.* 1991). The LFs were normalized vertically by matching the total counts in the range 17.7 < V < 18.7 (shown by the two vertical lines), and shifted horizontally to match the apparent mean V magnitudes of the horizontal branches. The blue HB stars in NGC 7492 were counted in the next brighter bin. The small metallicity differences between these objects (about 0.5 dex) are partially mitigated by matching the HB brightnesses. The horizontal shifts are consistent with the differences in the apparent distance moduli of the objects.

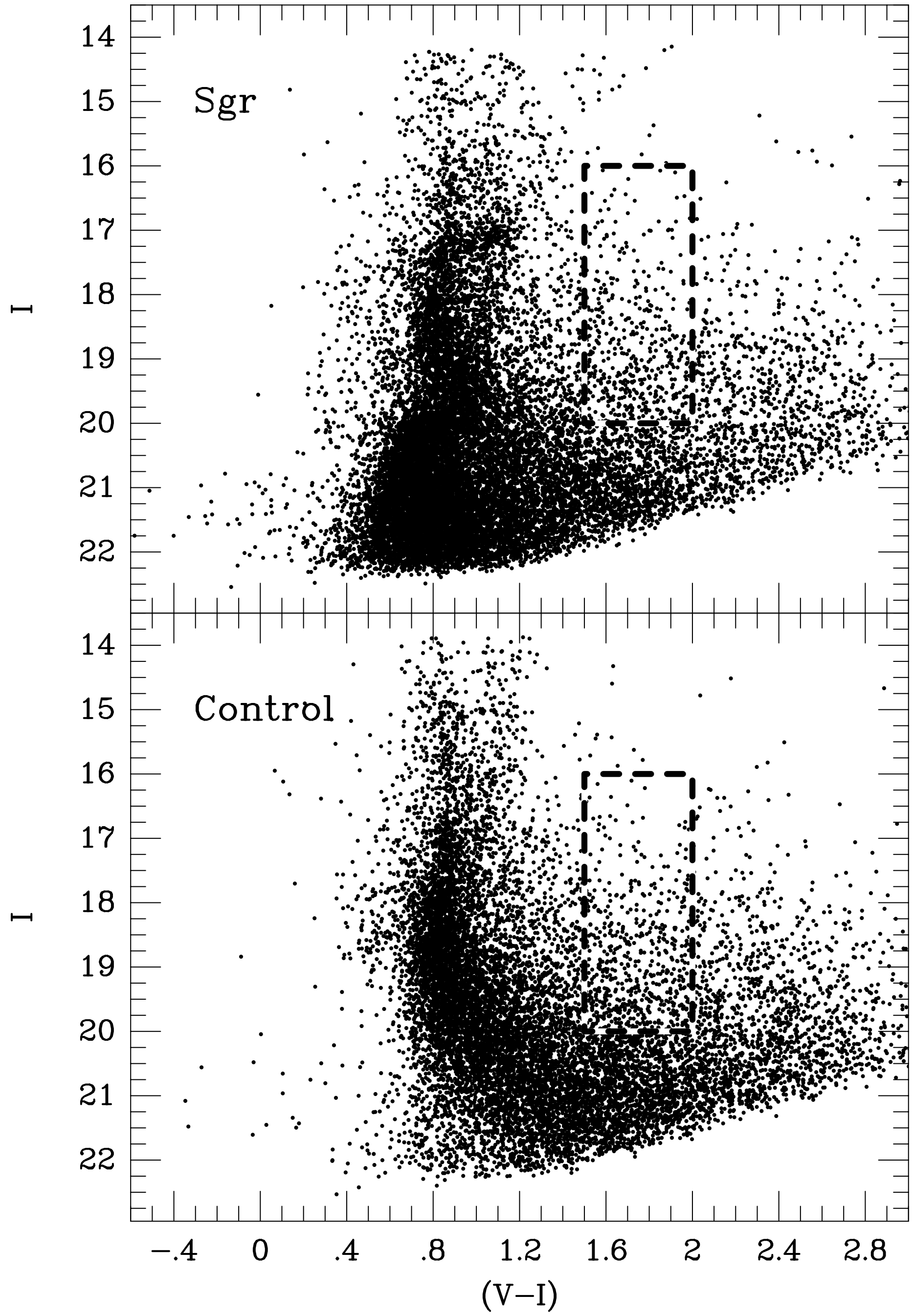

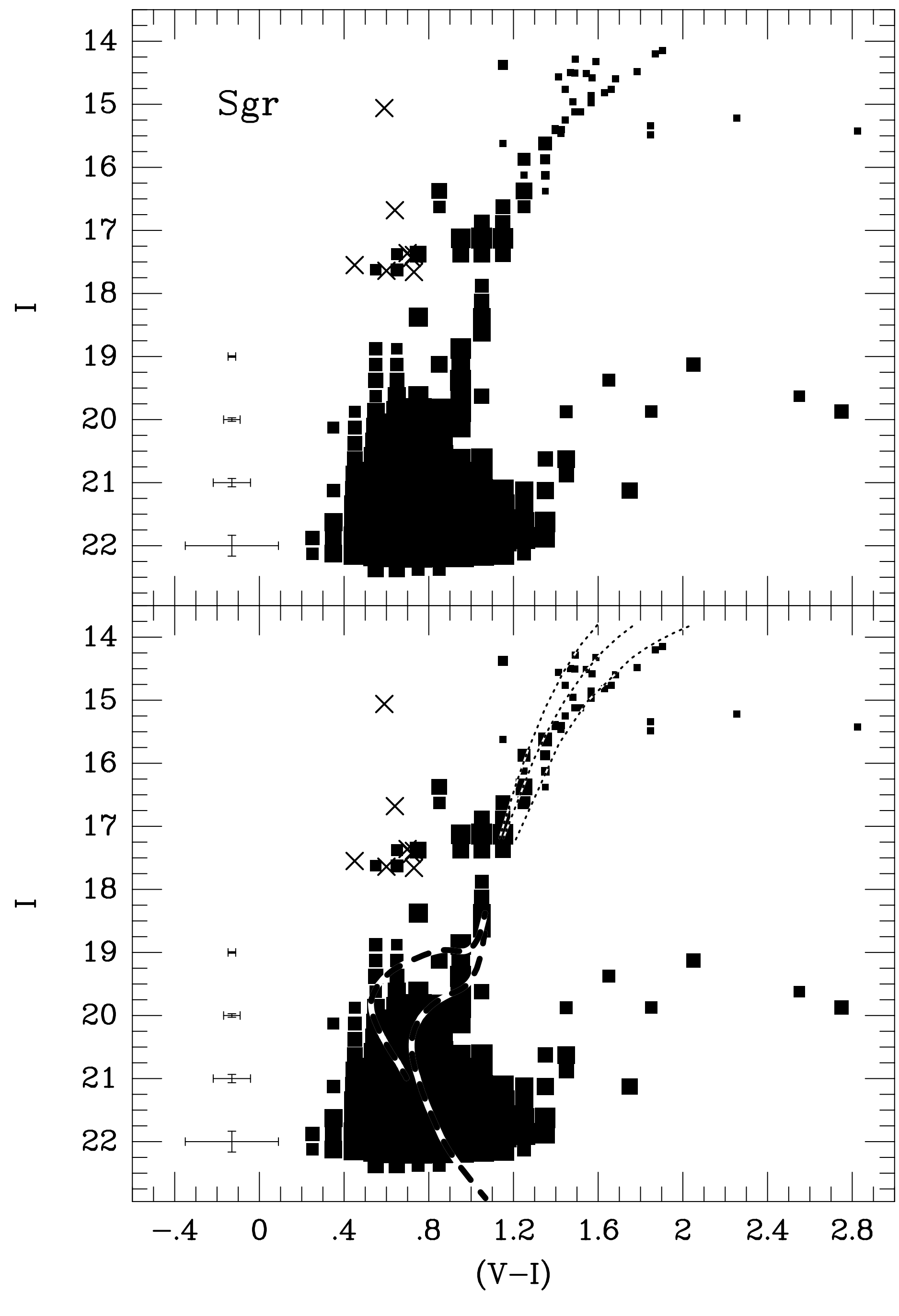

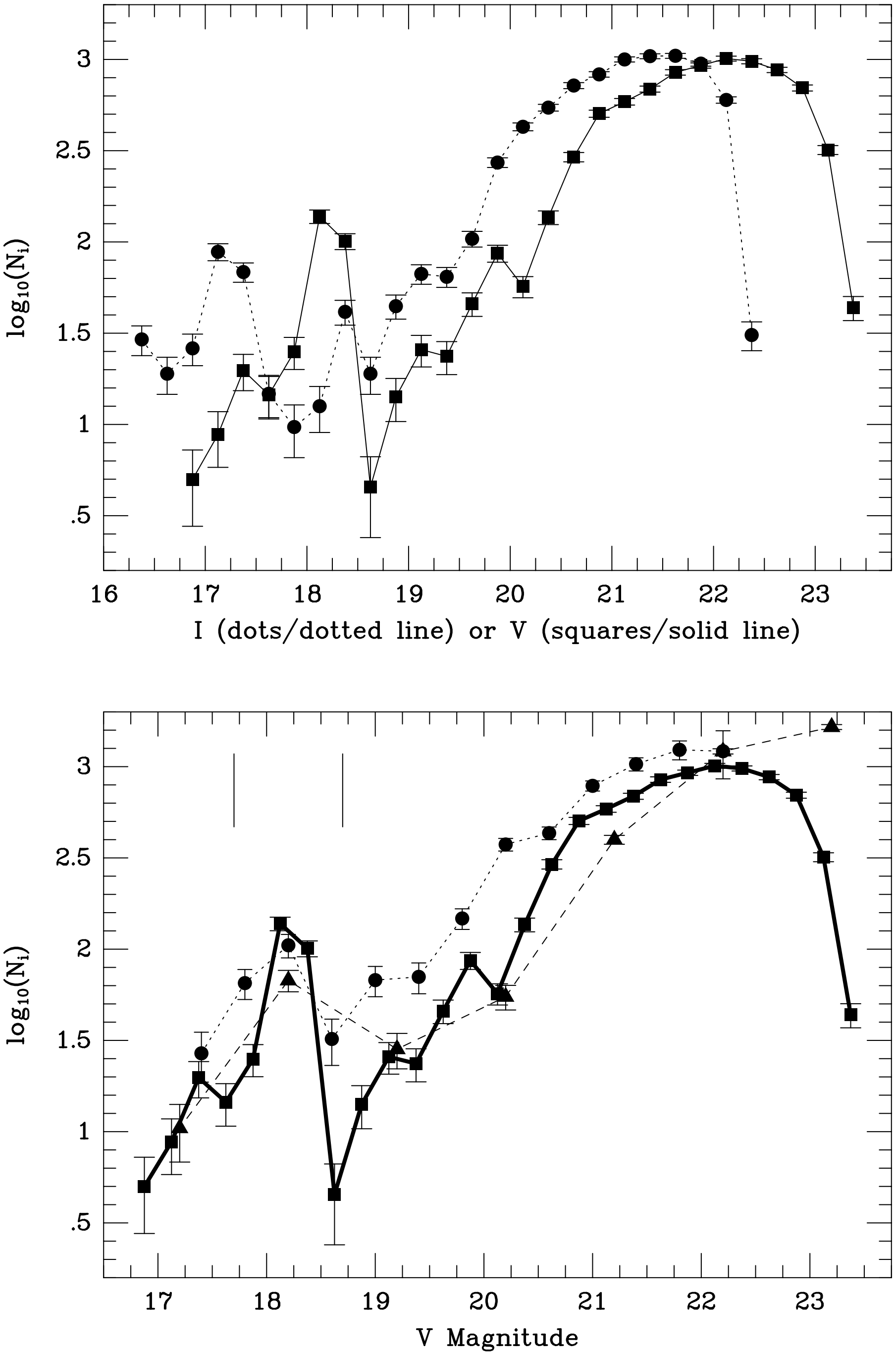